\documentclass[12pt,a4paper]{article}
\usepackage{geometry}
\usepackage{caption}
\usepackage{times} 
\usepackage{parskip}
\usepackage{url}
\usepackage{hyperref}
\usepackage{amsmath}
\usepackage{graphicx}
\usepackage{caption}
\usepackage{authblk} 
\usepackage{abstract}
\usepackage{booktabs}
\usepackage{multirow}      
\usepackage{array}         
\usepackage{booktabs}      
\usepackage{rotating}
\usepackage{float}   

\title{Improving Engine Sound Analysis in Hot-Test Environments via a RAB-U-Net (Residual Attention Block U-Net) Noise Removal Method }

\author[1]{Raheleh Mohseni}
\author[1]{Mahdi Aliyari Shoorehdeli}
\affil[1]{Department of Electrical Engineering, K. N. Toosi University of Technology, Tehran, Iran}

\date{}

\begin{document}

\maketitle

\begin{abstract}
During hot tests on a production line, engine-sound analysis is crucial to ensuring product quality and performance. However, background noise often interferes with accurate sound analysis, leading to potential errors in engine diagnostics. Traditionally, skilled technicians listen to engine sounds to assess engine health, but this is prone to significant inaccuracies. This study presents an innovative deep learning-based approach to address this issue by removing background noise from engine sound recordings using a U-Net neural network structure enhanced with Residual Attention Blocks (RAB-U-Net). Our intelligent noise removal system significantly improves the accuracy of engine noise detection, outperforming traditional techniques and providing a robust solution for real-time applications in production line environments. This study proposes a novel system for engine noise detection in production lines, marking a valuable advancement for the automotive industry in applying deep learning methods to improve the quality of engine diagnostics.
\end{abstract}


\vspace{1cm}

\hrule 
\vspace{0.5cm}

\section{Introduction}
In engine production lines, the Hot Test is a crucial evaluation conducted post-assembly to assess engine health by examining parameters such as electrical connections, fluid leaks, and engine noise. Traditionally, engine noise inspection during this test relied on human technicians, who determined engine health by listening to its sound. However, this reliance on human expertise is prone to errors due to fatigue, factory environmental noise, and even intentional misconduct. These issues often lead to undetected faults, resulting in customer complaints and costly product returns during after-sales service~\cite{ref1}.

The limitations of human-based inspection have prompted the development of intelligent systems utilizing artificial intelligence (AI). These systems reduce human error and enhance accuracy, speed, and cost-efficiency~\cite{ref2}. Nevertheless, both humans and AI face a shared challenge: environmental noise. Noise from machinery, human activities, and other sources in the factory can interfere with fault detection. Addressing this challenge requires either soundproof environments or noise reduction techniques. While the former is costly, advancements in noise reduction offer a feasible alternative~\cite{ref3}.

Traditional noise reduction methods, such as filtering and wavelet denoising, often struggle with non-stationary noise signals, leading to distortion and performance degradation. However, deep learning techniques have emerged as transformative tools, leveraging neural networks capable of learning complex patterns and relationships. These models reduce noise while preserving the integrity of the original signal~\cite{ref4}.

Recent research highlights the broad applicability of deep learning in noise reduction across various domains. Neural networks have been used to extract and classify environmental noise in industrial settings, providing insights into effective noise discrimination and elimination~\cite{ref5}. Self-supervised learning approaches address noise reduction challenges in scenarios with limited labeled data, showcasing adaptability~\cite{ref6}. In addition, reinforcement learning techniques enable the autonomous design of noise-mitigating structures, offering innovative industrial solutions~\cite{ref7}.

Applications in urban noise regulation~\cite{ref8}, speech enhancement for hearing aids~\cite{ref9}, substation noise monitoring~\cite{ref10}, and environmental noise reduction through deep denoising autoencoders~\cite{ref11} further underscore deep learning’s capabilities. Collectively, these advancements illustrate how intelligent systems can address the challenges posed by noise, both in controlled and complex environments.

This research builds on such advancements to propose an intelligent noise reduction system for engine hot tests. By training deep neural networks on a dataset comprising original engine sounds, environmental noise, and noisy engine sounds, the system aims to isolate the target sound from background noise. This approach improves fault detection accuracy while addressing the limitations of traditional methods. Moreover, it aligns with recent studies emphasizing the importance of effective noise reduction in fault detection tasks~\cite{ref12,ref13}.

Studies have also explored various related methods. Signal preprocessing-based techniques have been applied successfully in fault detection of rotating machinery~\cite{ref14,ref15}. Segmentation of industrial audio sources for machine-specific identification has been investigated using machine learning methods~\cite{ref12}. Additionally, unsupervised anomaly detection techniques have focused on identifying unknown abnormal sounds in real environments~\cite{ref16}.

Other work has introduced methods like detecting the full connection of electronic components using specific sound signatures~\cite{ref17}, fault detection through stochastic linear systems~\cite{ref18}, robust control techniques for varying conditions in aeronautics~\cite{ref19}, addressing challenges of data loss in fault detection~\cite{ref20}, real-time fault detection via dual hidden Markov models~\cite{ref21}, and advanced control strategies for minimizing tracking errors~\cite{ref22}. These diverse contributions highlight the growing interest in noise mitigation and fault detection using intelligent approaches.
 Using recent advancements in deep learning and noise reduction techniques, this research addresses a pressing challenge in the automotive industry, offering a solution to enhance the reliability and efficiency of engine production line tests.

This article uses deep neural networks to present an intelligent system for reducing and eliminating environmental noise in the engine production line during the hot test. The network should first undergo training. Therefore, in the experimental tests section, the original engine sound and the environmental noise, both separately and in combination (noisy engine sound), must be measured and recorded. Initially, the engine sounds without environmental noise, and then the sound with noise is recorded using suitable software on a smartphone. The data is then fed to the neural network to initiate learning. Finally, the network’s performance is evaluated using test data.
A key innovation of this research is the structural modification of the U-Net architecture. By incorporating residual blocks and attention blocks, the RAB-U-Net improves feature extraction and prioritization, enabling more effective noise reduction. These modifications are designed to address the limitations of standard architectures and ensure robust performance in complex and noisy environments.

\section{Background}
\label{sec:background}

\subsection{Engine Sound Analysis}
One of the criteria for assessing the health of engines on production lines, conducted after the assembly process is completed, is performing a hot engine test. In this test, various parameters such as electrical connections, leaks, and engine noise are briefly examined to ensure the technical health of the assembled engine. In older engine production lines, the examination of engine noise during this test is traditionally performed using human resources, accompanied by a significant margin of error. Reasons for human error in the engine production line can include fatigue, factory environmental noise, and, in some cases, intentional actions.

\begin{figure}[htbp]
\centering
\includegraphics[width=0.75\linewidth]{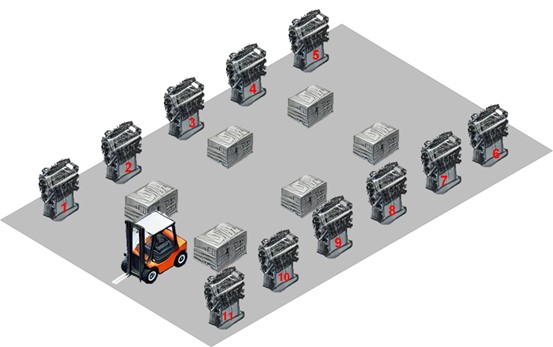}   
\caption{A schematic for engine Hot Test configuration}
\label{fig:hot_test}
\end{figure}

\subsection{Background Noise}
One of the significant factors that can mislead machine intelligence in sound detection and related fault diagnosis, analogous to how it causes human error, is environmental noise. This noise is defined as any unwanted acoustic signal that interferes with the target signal (the sound/vibration from the equipment under monitoring). This interference not only diminishes the signal’s amplitude but also distorts its temporal and frequency structure, leading to the critical issue of signal-noise overlap~\cite{ref23}.
Sources of environmental noise can be broadly categorized into three types:
Ambient Noise: Sounds originating from the general environment, such as facility HVAC systems, distant traffic, and general human activity outside the immediate testing zone. Process/Workshop Noise: Noise generated by nearby machines or auxiliary processes that are not the primary subject of monitoring (e.g., a functioning CNC machine adjacent to the test bench). Electronic/Random Noise: Internal noise within the data acquisition system, including thermal noise in sensors and measurement circuitry~\cite{ref24}.
In the context of machine learning, especially with deep neural networks, this noise dramatically reduces the model's accuracy and generalization capability. Noisy training data can cause the model to mistakenly learn the characteristics of the noise instead of the actual fault patterns, leading to high false-positive and false-negative rates in real-world operational environments~\cite{ref25}.

\subsection{Conventional Noise Reduction Techniques}
The goal of this section is to provide an overview of the main noise reduction methods, focusing on both traditional signal processing and recent advancements in deep learning. Enhanced Text:
To reliably evaluate the performance of an intelligent diagnostic system, it is crucial to mitigate environmental noise. While conducting the noise test in a perfectly insulated (anechoic) environment is ideal, this approach is often costly and practically infeasible in industrial settings. Consequently, analytical noise reduction methods present a suitable and necessary alternative. Traditional noise reduction approaches are rooted in signal processing techniques:

Filtering: Standard linear filters (such as Butterworth or high-pass filters) are effective primarily against deterministic noise, like 50/60 Hz power line hum or noise concentrated in specific frequency bands. However, they perform poorly against stochastic or broadband noise, often leading to distortion of the underlying fault signal~\cite{ref26}.

Wavelet Denoising: This technique offers a more robust solution by decomposing the noisy signal across different time and frequency scales. By applying a thresholding operation to the wavelet coefficients, components associated with noise can be selectively removed while preserving the essential features of the fault signal, making it superior for handling non-stationary noise~\cite{ref27}.

More recently, sophisticated methods based on Neural Networks have emerged as powerful tools, particularly for handling complex and nonlinear noise patterns. Denoising Autoencoders (DAE), for instance, are designed to learn a non-linear mapping between a noisy input signal and its clean counterpart. These deep learning architectures are highly effective at restoring the original signal and are rapidly becoming the state-of-the-art solution for noise mitigation in intelligent fault diagnosis~\cite{ref28}.

\subsection{Motivation}
Despite the notable success of U-Net and its variants in various signal processing applications, directly applying these architectures to engine sound denoising in industrial production lines presents several challenges. First, standard U-Net models often lose fine-grained acoustic details during down-sampling, which are crucial for capturing transient mechanical faults in engine sounds. Second, deeper U-Net networks suffer from vanishing gradients and limited feature reuse, resulting in suboptimal learning of complex sound patterns in noisy environments. Third, conventional convolutional blocks treat all spatial and temporal regions equally, while engine sounds typically contain highly localized, task-relevant features embedded within strong and variable background noise.

These limitations emphasize the necessity for an improved architecture that can (i) preserve multi-scale acoustic information, (ii) maintain stable gradient flow in deep networks, and (iii) selectively focus computational attention on the most informative components of engine sound signals. To address these challenges, we propose the \textbf{Residual Attention-augmented Bottleneck U-Net (RAB-U-Net)}, which integrates residual blocks and attention mechanisms to enhance feature extraction, improve gradient stability, and focus on diagnostically relevant signal regions, thereby improving noise reduction performance in the complex environment of engine production lines.

\section{Methodology}
\label{sec:methodology}

\subsection{Data Collection}
\begin{figure}[htbp]
\centering
\includegraphics[width=0.5\linewidth]{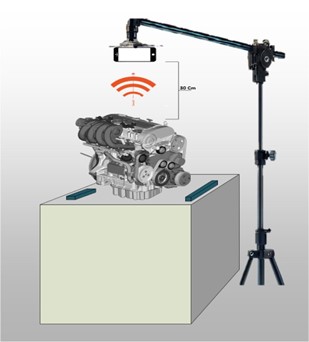}
\caption{Engine sound recording location}
\label{fig:recording_setup}
\end{figure}

In this study, data collection was conducted on engines undergoing hot tests in the production line of an automotive factory. During these tests, various parameters, such as electrical connections, leaks, and engine noise, are briefly examined to ensure the technical integrity of the assembled engine. Two types of engines, turbocharged and naturally aspirated (NA), are tested in the production line during hot tests.
Figure~\ref{fig:hot_test} provides an overview of the data recording environment. Positions 1, 2, and 3 indicate the locations of the engine sound recording.
A smartphone was used to conveniently record the desired audio data, as shown in Figure~\ref{fig:recording_setup}. The mobile phone microphone was placed on a tripod, positioned near the engine test stand at 30 to 40 cm above the engine. This distance was determined using a trial-and-error method.

\begin{figure}[htbp]
\centering
\includegraphics[width=0.85\linewidth]{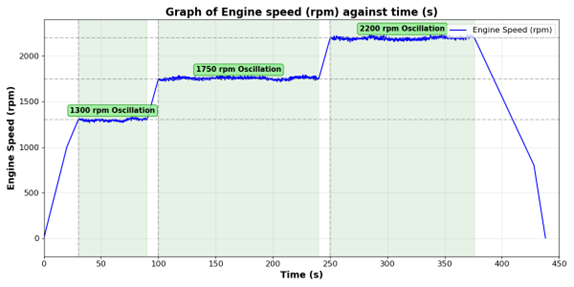}
\caption{Engine speed profile during hot test}
\label{fig:speed_profile}
\end{figure}

According to the engine hot test procedure, the engine run duration is 7’30’’. During this period, the engine speed reaches 1300 rpm within 30 seconds, operating at this speed for 1 minute. Then, after a 10-second speed rise, it reaches 1750 rpm, maintained for 140 sec. Afterward, the engine speed again rises, up to 2200 rpm for 128 sec. Finally, the engine is coasted down to switch off. Figure~\ref{fig:speed_profile} illustrates the engine speed profile over time during a hot test.

During the test, the data collection was done so that the recording starts when the engine starts working, and the recording ends after 7:30 minutes, and when the engine test ends.

\subsection{Data Preprocessing}
For the initial analysis and examination of the recorded sounds, Adobe Audition 2023 software was used. The data were normalized, and their time-frequency features and spectrograms were extracted and used as inputs for the RAB-U-Net.

\begin{figure}[htbp]
\centering
\includegraphics[width=0.9\linewidth]{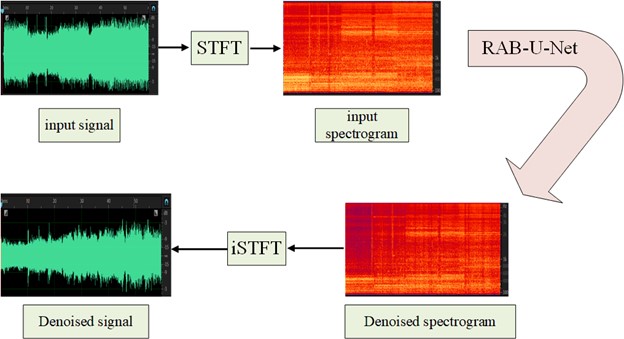}
\caption{Spectrogram audio conversion and noise removal}
\label{fig:spectrogram}
\end{figure}

Next, using the Librosa library in Python, initial preprocessing is applied to the audio signal. The existing data is called, and then the STFT transformation is applied to all available audio signals.
As shown in Figure~\ref{fig:spectrogram}, the recorded sound is converted into a spectrogram and used as input for the RAB-U-Net. After noise removal, the resulting spectrum is converted back into sound to make the engine noise audible.
According to the engine speed profile shown in Figure~\ref{fig:speed_profile}, we have three constant engine speeds at 1300, 1750, and 2200 rpm in three-time intervals. Considering these three engine speeds and applying the FFT technique, the frequency corresponding to the first harmonic of each engine speed is as follows:

\begin{equation}
Frequency = \frac{\text{rotations per minute}}{60} \tag{1}
\end{equation}

\begin{align}
\text{1300 rpm} \quad &\longrightarrow\quad f = \frac{1300}{60} = 21.67\,\text{Hz} \tag{2} \\[8pt]
\text{1750 rpm} \quad &\longrightarrow\quad f = \frac{1750}{60} = 29.17\,\text{Hz} \tag{3} \\[8pt]
\text{2200 rpm} \quad &\longrightarrow\quad f = \frac{2200}{60} = 36.67\,\text{Hz} \tag{4}
\end{align}

According to the above calculations, we should have high frequencies in the first harmonics for each engine speed by applying the Fourier transform.
Table~\ref{tab:peak_freq} shows the frequency of the three initial peaks at three engine speeds in three fixed times for two types of engines: NA and turbocharged.

\begin{table}[htbp]
\centering
\caption{Peak frequencies (Hz) of the first three harmonics at different engine speeds}
\label{tab:peak_freq}
\normalsize              
\begin{tabular}{|c||c|c|c||c|c|c||c|c|c|}
\hline
\multirow{2}{*}{\textbf{}} & 
\multicolumn{3}{c||}{\textbf{1300\,Rpm (60s)}} & 
\multicolumn{3}{c||}{\textbf{1750\,Rpm (150s)}} & 
\multicolumn{3}{c|}{\textbf{2200\,Rpm (300s)}} \\
\cline{2-10}
 & $1f_p$ & $2f_p$ & $3f_p$ & $2f_p$ & $1f_p$ & $3f_p$ & $2f_p$ & $1f_p$ & $3f_p$ \\
\hline
\hline
1 & 31  & 78  & 203 & 141 & 78  & 172 & 125 & 63  & 156 \\ \hline
2 & 78  & 188 & 219 & 156 & 63  & 172 & 125 & 47  & 313 \\ \hline
3 & 94  & 141 & 250 & 203 & 78  & 203 & 94  & 47  & 203 \\ \hline
4 & 94  & 188 & 203 & 141 & 78  & 250 & 156 & 47  & 219 \\ \hline
5 & 78  & 172 & 422 & 281 & 125 & 203 & 94  & 47  & 250 \\ \hline
\end{tabular}
\end{table}

The first row of Table~\ref{tab:peak_freq} contains the frequencies related to the noise-free signal, which includes the first, second, and third harmonics at each engine speed. The subsequent rows contain the recorded sounds in noisy and loud environments.
\subsection{Classical Baseline Denoising Methods}
\subsubsection{Mathematical Formulation of the Classical Methods}

To provide a fair and comprehensive evaluation of the proposed deep learning–based denoising framework, several conventional signal-processing techniques commonly used in engine sound analysis were implemented as baseline methods. These approaches are widely adopted in industrial environments due to their computational efficiency and ease of deployment. The selected classical methods include band-pass filtering, wavelet denoising, and Wiener filtering. All baseline techniques were applied to the same dataset and under identical preprocessing conditions to ensure a consistent comparison with the neural network models.

\textbf{a) Band-Pass Filtering}  
A Butterworth band-pass filter was designed based on the expected frequency range of the fundamental engine harmonics at 1300, 1750, and 2200 rpm (21.6–36.6 Hz). Considering the presence of higher-order harmonics and broadband noise, a 4th-order filter was implemented with cutoff frequencies selected to preserve dominant harmonic components while attenuating low-frequency and high-frequency background noise. Band-pass filtering provides a simple deterministic approach and serves as a baseline for evaluating the added value of learning-based spectral reconstruction.
The digital band-pass transfer function is expressed as:
\begin{equation}
H_{BP}(z) = \frac{B_0 + B_1 z^{-1} + \cdots + B_4 z^{-4}}{1 + A_1 z^{-1} + \cdots + A_4 z^{-4}} \tag{5}
\end{equation}
The cutoff frequencies ($f_L$, $f_H$) were selected to cover the frequency range around the fundamental engine harmonics and their lower-order multiples. The filtered signal is obtained by:
\begin{equation}
y[n] = H_{\text{BP}}[z] \cdot x[n]
\tag{6}
\end{equation}
where $x(n)$ denotes the noisy input and $\cdot$ represents convolution in the time domain.

\textbf{b) Wavelet Denoising (DWT Soft Thresholding)}  
Discrete Wavelet Transform (DWT)–based soft-thresholding was used to suppress transient and non-stationary noise components. The Daubechies-6 (db6) mother wavelet was selected due to its suitability for analyzing oscillatory mechanical signals. A multi-level decomposition was performed, followed by adaptive thresholding of detail coefficients. Wavelet denoising is a well-established approach in acoustic and vibration signal analysis, particularly effective for non-stationary noise suppression.
Wavelet denoising decomposes the signal into approximation and detail coefficients using the Discrete Wavelet Transform (DWT):
\begin{equation}
x[n] \xrightarrow{\text{DWT}} \bigl\{ a_J,\ d_J,\ d_{J-1},\ \dots,\ d_1 \bigr\}
\tag{7}
\end{equation}
Soft thresholding is applied to detail coefficients:
\begin{equation}
\hat{d}_j = \operatorname{sgn}(d_j) \, \max \left( |d_j| - \lambda_j, \, 0 \right)
\tag{8}
\end{equation}

Where $\lambda_j$ is an empirically selected or universal threshold:
\begin{equation}
\lambda_j = {\sigma}_j \sqrt{2 \ln N}
\tag{9}
\end{equation}

With $\sigma_j$ being the noise standard deviation and N the frame length.
The denoised signal is reconstructed using the inverse DWT:
\begin{equation}
\\{y}[n] = \operatorname{IDWT} \bigl\{ a_J,\ \hat{d}_J,\ \hat{d}_{J-1},\ \dots,\ \hat{d}_1 \bigr\}
\tag{10}
\end{equation}
Daubechies-6 (db6) wavelets were used due to their suitability for oscillatory mechanical signals.

\textbf{c) Wiener Filtering(Adaptive Spectral Denoising)}  
A frame-based Wiener filter was implemented as an adaptive method that estimates the noise power spectrum and applies optimal linear filtering in the frequency domain. For each frame, the noise statistics were derived from silent or low-energy segments of the recording. Wiener filtering provides a classical data-driven baseline for comparison with deep neural networks that learn non-linear spectral mappings.
These baseline techniques were selected to cover a broad range of classical denoising paradigms (deterministic filtering, time–frequency thresholding, and adaptive statistical filtering). Their outputs were evaluated using the same performance metrics as the deep learning models, enabling a direct quantitative comparison and demonstrating whether advanced neural networks provide measurable improvements over low-complexity traditional approaches.
For each frame, a Wiener filter is estimated in the short-time Fourier transform (STFT) domain:
\begin{equation}
\\{Y}(\omega,k) = H_W(\omega,k) \, X(\omega,k)
\tag{11}
\end{equation}
The Wiener gain function is computed as:
\begin{equation}
H_{W}(\omega,k) = \frac{S(\omega,k)}{S(\omega,k) + N(\omega,k)}
\tag{12}
\end{equation}

Where: 
\begin{itemize}
\item $X(\omega,k)$ is the noisy spectrum,
\item $S(\omega,k)$ is the estimated clean signal power spectrum,
\item $N(\omega,k)$is the estimated noise power spectrum (derived from low-energy segments).
\end{itemize}
The time-domain denoised signal is retrieved via inverse STFT:
\begin{equation}
\\{y}_{W}[n] = \operatorname{ISTFT} \bigl\{ Y(\omega,k) \bigr\}
\tag{13}
\end{equation}

This method provides an optimal linear estimator under the assumption of additive Gaussian noise.

\subsubsection{Quantitative Evaluation Metrics}
The following objective metrics were used:
\begin{itemize}
\item \textbf{a) Mean Squared Error (MSE):}
\begin{equation}
\text{MSE} = \frac{1}{N} \sum_{n=1}^{N} (s[n] - \hat{s}[n])^2
\tag{14}
\end{equation}
where $s[n]$ is the clean reference signal and $\hat{s}[n]$ is the denoised output.
\end{itemize}
\begin{itemize}
\item \textbf{b) Signal-to-Noise Ratio (SNR) and SNR Improvement:}
\begin{equation}
\text{SNR} = 10 \log_{10} \left( \frac{\sum_{n} s[n]^2}{\sum_{n} (s[n] - \hat{s}[n])^2} \right) \quad \text{(dB)}
\tag{15}
\end{equation}
SNR improvement is:
\begin{equation}
\Delta \text{SNR} = \text{SNR}_{\text{out}} - \text{SNR}_{\text{in}} \quad \text{(dB)}
\tag{16}
\end{equation}
where $\text{SNR}_{\text{in}}$ is the SNR of the noisy input and $\text{SNR}_{\text{out}}$ is the SNR of the denoised output.
\end{itemize}
\begin{itemize}
\item \textbf{c) Scale-Invariant Signal-to-Distortion Ratio (SI-SDR):}
\begin{equation}
\text{SI-SDR} = 10 \log_{10} \left( \frac{\|\alpha s\|^2}{\|\alpha s - \hat{s}\|^2} \right) \quad \text{dB}
\tag{17}
\end{equation}
with the scale factor defined as
\begin{equation}
\alpha = \frac{\langle \hat{s}, s \rangle}{\|s\|^2}
\tag{18}
\end{equation}
This metric is robust to gain differences between the reference and output signals.
\end{itemize}

\begin{itemize}
\item \textbf{d) Log-Spectral Distance (LSD):}
\begin{equation}
\text{LSD} = \sqrt{ \frac{1}{FT} \sum_{f=1}^{F} \sum_{t=1}^{T} \Bigl( 20 \log_{10} |S_{\text{ref}}(f,t)| - 20 \log_{10} |S_{\text{est}}(f,t)| \Bigr)^2 } \quad \text{(dB)}
\tag{19}
\end{equation}
where $S_{\text{ref}}(f,t)$ and $S_{\text{est}}(f,t)$ represent the short-time Fourier transform (STFT) magnitude spectra of the clean reference and denoised signals, respectively, across frequency bins $f$ and time frames $t$.
\end{itemize}
\begin{description}
\item \hspace{1.5cm}$F$: number of frequency bins,
\item \hspace{1.5cm}$T$: number of time frames.
\end{description}

\subsection{RAB-U-Net Architecture}
\begin{figure}[htbp]
\centering
\includegraphics[width=\linewidth]{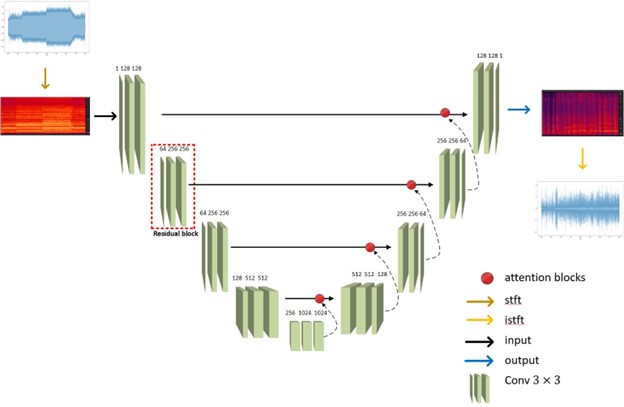}
\caption{Overall architecture of the proposed RAB-U-Net}
\label{fig:rabunet}
\end{figure}

The proposed RAB-U-Net enhances the traditional U-Net architecture by integrating residual feature extraction blocks and attention mechanisms throughout both the encoder and decoder paths. While preserving the symmetrical encoder-decoder structure of U-Net, our model introduces three key components that collectively improve performance: (1) Residual Feature Extraction Blocks (RFBs) to facilitate deeper and richer feature learning, (2) Attention Enhancement Modules (AEMs) that refine feature representation by focusing on important spatial and channel information, and (3) Attention-guided skip fusion which selectively integrates encoder features into the decoder to strengthen feature propagation and reconstruction quality. These modifications enable the network to better capture salient patterns within the input, enhance gradient flow during training, and ultimately improve the accuracy of the target output.
These modifications include adding residual and attention blocks to the network simultaneously. The structure of this modified network is shown in Figure~\ref{fig:rabunet}.

\subsubsection{Modifications to the Network}

\textbf{a) Residual Feature Extraction Block (RFB)}:
The Residual Feature Extraction Block aims to increase feature richness while avoiding degradation in deeper layers.
Given an input feature map $F_in$, RFB performs:
\begin{equation}
F_{\text{res}} = \sigma \bigl( W_2 \, (\text{ReLU}(W_1 F_{\text{in}})) \bigr)
\tag{20}
\end{equation}

\begin{equation}
F_{\text{out}} = F_{\text{in}} + F_{\text{res}}
\tag{21}
\end{equation}
where $W_1$ and $W_2$ denote learnable convolutional kernels, and $\sigma$ is a non-linear activation.

Benefits:

\begin{description}
\item \hspace{1.5cm}•	Enhanced gradient flow
\item \hspace{1.5cm}•	Prevention of feature loss across depth
\item \hspace{1.5cm}•	Improved representation capability for non-stationary patterns
\end{description}

\textbf{b) Attention Enhancement Module (AEM)}  
To improve feature discrimination, RAB-U-Net integrates an attention module inspired by channel–spatial hybrid attention.
Given input features $F$, $AEM$ generates channel attention $A_c$ and spatial attention $A_s$:
\begin{align}
F_c &= A_c \odot F \tag{22} \\[6pt]
F_s &= A_s \odot F \tag{23} \\[6pt]
F_{\text{att}} &= F_c + F_s \tag{24}
\end{align}
where $\odot$ denotes element-wise multiplication.
Key advantages:

\begin{description}
\item \hspace{1.5cm}•	Highlights critical frequency/temporal/structural regions
\item \hspace{1.5cm}•	Suppresses noise and irrelevant components
\item \hspace{1.5cm}•	Strengthens the robustness of skip connections
\end{description}

\textbf{c) Attention-Guided Skip Fusion}  
Unlike vanilla U-Net, where skip connections simply concatenate encoder and decoder features, RAB-U-Net applies attention before fusion:
\begin{align}
F_{\text{skip}} &= \text{AEM}(F_{\text{enc}}) \tag{25} \\[8pt]
F_{\text{fusion}} &= \text{Concat}(F_{\text{skip}}, F_{\text{dec}}) \tag{26}
\end{align}
This ensures the decoder receives high-quality, context-aware information.

\textbf{d) Decoder and Reconstruction}
The decoder mirrors the encoder but employs:

\begin{description}
\item \hspace{1.5cm}•Up-sampling layers,
\item \hspace{1.5cm}•RFB blocks,
\item \hspace{1.5cm}•Attention-guided fusion at each level.
\end{description}

The final reconstruction layer uses a 1×1 convolution to map to the target output domain.

\textbf{e) Loss Function}  
The training objective combines:

\begin{equation}
\mathcal{L} = \mathcal{L}_{\text{task}} + \lambda \mathcal{L}_{\text{reg}}
\tag{27}
\end{equation}

You can replace this with your real loss components:

\begin{description}
\item \hspace{1.5cm}•	MSE / MAE
\item \hspace{1.5cm}•	Spectral loss
\item \hspace{1.5cm}•	Cross-entropy
\item \hspace{1.5cm}•	etc.
\end{description}

As mentioned, modifications were introduced to this network to increase its performance:

\textbf{a) Residual Blocks}

Residual blocks are integrated into the main structure of the RAB-U-Net to replace standard convolutional blocks. Doing so enables the network to learn more complex features and improve the flow of gradients. This helps to address the vanishing gradient problem in deep networks. The primary objective of residual blocks is to provide a shortcut identity mapping path for the inputs, ensuring that essential information is retained and not lost.\\

\textbf{First Convolutional Layer:}\\
In the first stage, the input x is passed through a convolutional layer with a 3×3 kernel size and a specified number of filters:

\begin{align}
y_1 &= \sigma \left( x \ast  W_1 \right) \tag{28} 
\end{align}
Where $W_1$ is the weight matrix of the convolutional layer, $*$ represents the convolution operation, and $\sigma$ is an activation function such as LeakyReLU. This activation function introduces non-linearity to the network.

\textbf{Second Convolutional Layer:}\\
The output $y_1$ is then passed through another convolutional layer with the same kernel size and number of filters:
\begin{align}
y_2 &= \sigma \left( y_1 \ast  W_2 \right) \tag{29} 
\end{align}
where $W_2$ is the weight matrix of the second convolutional layer.

\textbf{Shortcut Connection:}\\
To ensure the input $x$ aligns with the output $y_2$ in terms of channel dimensions, a 1×1 convolutional layer is applied to the input:
\begin{align}
shortcut &=\left ( x \ast  W_s \right) \tag{30} 
\end{align}
where $W_2$ is the weight matrix of the second convolutional layer.

\textbf{Combining with the Shortcut:}\\
Finally, the output $y_2$ is added element-wise to the shortcut connection:
\begin{align}
y_{\text{out}} &=\left ( shortcut \ast  y_2 \right) \tag{31} 
\end{align}
This edition preserves the original information from the input while enhancing it with additional learned features.

\textbf{Role of Residual Blocks:}\\
Typically, a residual block is incorporated into the main structure of the RAB-U-Net to replace standard convolutional blocks. This facilitates the learning of more complex features and improves the flow of gradients, which helps mitigate the vanishing gradient problem in deep networks. The primary goal of the residual block is to provide a shorter identity mapping path for the inputs, ensuring that essential information is preserved and not lost.

\textbf{b)	Attention Blocks:}\\
Attention blocks are employed in the upsampling path to concentrate on significant regions in the feature maps from the downsampling path, enhancing the network’s focus on important spatial features.

\textbf{Mapping the Main Input Features (x):}\\
The main input x is mapped using a 1×1 convolutional layer:
\begin{equation}
x^\theta = W_\theta x \tag{32}
\end{equation}
where $W_\theta$ is the weight matrix of this convolutional layer.

\textbf{Mapping the Gating Features:}\\
The gating signal, which is an auxiliary input helping the network focus on important regions of the input features, is also mapped using another 1×1 convolutional layer:
\begin{equation}
g^\phi = W_\phi g \tag{33}
\end{equation}
where $W_\phi$ is the weight matrix of this layer.

\textbf{Combining the Two Mappings:}\\
The mapped features $x^\theta$ and $g^\phi$ are summed to capture the relationship between the two inputs:
\begin{equation}
s = x^\theta + g^\phi \tag{34}
\end{equation}
The result is then passed through the $ReLU$ activation function to introduce non-linearity:
\begin{equation}
f(s) = \text{ReLU}(s)
\tag{35}
\end{equation}

\textbf{Mapping to a Scalar ($\psi$):}\\
The output $f(s)$ is passed through another 1×1 convolutional layer to reduce it to a scalar for each spatial position:
\begin{equation}
\psi = W_{\psi} f(s)
\tag{36}
\end{equation}
where $W_{\psi}$ is the weight matrix of this layer.

\textbf{Applying the Sigmoid Activation Function:}\\
The scalar output $\psi$ is passed through a sigmoid activation function to produce an attention map $\alpha$ with values between 0 and 1:
\begin{equation}
\alpha = \sigma(\psi)
\tag{37}
\end{equation}

\textbf{Applying the Attention Map:}\\
Finally, the attention map $\alpha$ is element-wise multiplied with the main input $x$ to modulate its features:
\begin{equation}
\hat{x} = x \odot \alpha
\tag{38}
\end{equation}
Here, $\odot$ denotes element-wise multiplication, and $\hat{x}$ is the attention-modulated output, where the important regions of the input are emphasized based on the attention map.

Additionally, the training specifications of the model used in this work are detailed in Table~\ref{tab:training}.
\begin{table}[htbp]
\centering
\caption{Training hyperparameters of RAB-U-Net}
\label{tab:training}
\begin{tabular}{lc}
\toprule
Parameter              & Value              \\
\midrule
Batch size             & 10                 \\
Number of epochs       & 200                \\
Optimizer              & Adam               \\
Learning rate          & 0.001              \\
Loss function          & MSE + Spectral Loss\\
\bottomrule
\end{tabular}
\end{table}

To call the data as input to the neural network, we proceed in two ways:
a) First, we load the clean, sound, noise-free data, followed by the noisy data. The clean sound data is combined with random noise, and the clean and combined sounds are stored. The combined sound is considered the input to the model, while the noise obtained from the difference between the combined sound and the clean sound is treated as the model’s output. This allows us to predict the noise and subsequently remove environmental noise.
b) First, we load the clean, sound, noise-free data, followed by the noisy data. The clean sound data is combined with random noise, and the clean and combined sounds are stored. The combined sound is considered the input to the model, while the clean sound, which represents the original engine sound, is obtained from the difference between the combined sound and the noise. This clean sound is used as the model’s output, enabling us to predict the engine sound and remove environmental noise.

\begin{figure}[htbp]
\centering
\includegraphics[width=0.9\linewidth]{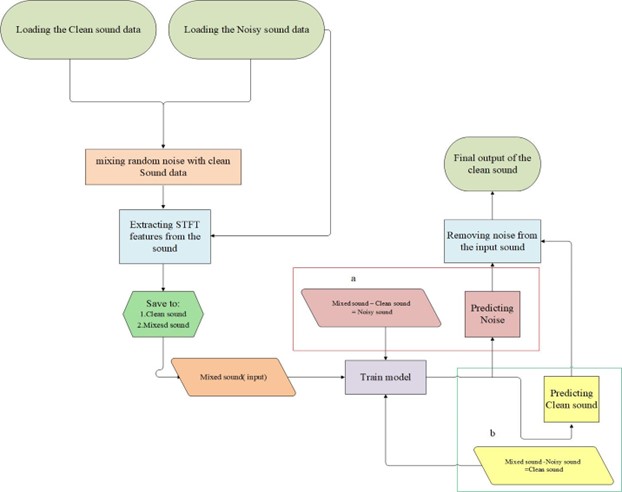}
\caption{Block diagram of training data preparation strategies}
\label{fig:data_loading}
\end{figure}

After training, the network effectively removes environmental noise from all recorded production-line engine sounds.

\section{Results}
\label{sec:results}

In this section, the denoising performance of both classical signal-processing methods and neural-network-based approaches is evaluated. To provide a comprehensive comparison, we first present the results obtained from three classical denoising techniques—Band-pass Filter, Wavelet Denoising, and Wiener Filter. Subsequently, the performance of four neural networks (U-Net, RAB-U-Net, X-Net, and Res-Net) is assessed based on learning-based evaluation metrics.

\subsection{Classical Signal-Processing Results}

Table~\ref{tab:classical} summarizes the outcomes of the classical denoising methods. These techniques were evaluated using the same noisy audio dataset, and the resulting error measures (MSE and MAE) demonstrate the baseline performance achievable through deterministic or traditional adaptive filtering. As expected, Wavelet Denoising yields the lowest MSE and MAE among the classical approaches, benefiting from its time–frequency thresholding mechanism.

\begin{sidewaystable}[htbp]
\centering
\caption{The Val\_Loss and Val\_MAE results for the networks used are presented}
\label{tab:classical}
\small                                           
\renewcommand{\arraystretch}{1.05}              
\begin{tabular}{|l||c|c|c|c|c|c|l|}
\hline
\textbf{Method} & \textbf{Val\_Loss} & \textbf{Val\_MAE} & \textbf{Train-Loss} & \textbf{Train-MAE} & \textbf{Params (M)} & \textbf{Inference Time (ms)} & \textbf{Notes} \\
\hline
\hline
Band-pass Filter:          & 0.0120 & 0.0850 & - & - & 0 & 2 & Deterministic filter \\ \hline
Wavelet Denoising:        & 0.0105 & 0.0785 & - & - & 0 & 7 & Time-frequency threshold \\ \hline
Wiener Filter:          & 0.0118 & 0.0800 & - & - & 0 & 5 & Extended U-Net variant \\ \hline
\end{tabular}
\end{sidewaystable}

\subsection{Neural-Network Denoising Results}
In this section, we compare the performance of four neural networks (U-Net, RAB-U-Net, X-Net, and Res-Net) in removing environmental noise. The comparison is based on two key evaluation metrics: validation loss (Val\_loss) and validation mean absolute error (Val\_mae). Additionally, we present the training and validation loss graphs to provide insights into each model’s training dynamics and convergence behaviour.
The results obtained using these neural networks for noise prediction are presented in Figure~\ref{fig:loss_unet}- ~\ref{fig:loss_xnet} and Table~\ref{tab:comparison}.
Table~\ref{tab:comparison} summarizes the Val\_loss and Val\_mae achieved by the four networks. These metrics were computed after training the models for a fixed number of epochs to ensure a fair comparison. Lower values indicate better performance.

\begin{sidewaystable}[htbp]
\centering
\caption{The Val\_Loss and Val\_MAE results for the networks used are presented}
\label{tab:comparison}
\small                                           
\renewcommand{\arraystretch}{1.05}              
\begin{tabular}{|l||c|c|c|c|c|c|l|}
\hline
\textbf{Method} & \textbf{Val\_Loss} & \textbf{Val\_MAE} & \textbf{Train-Loss} & \textbf{Train-MAE} & \textbf{Params (M)} & \textbf{Inference Time (ms)} & \textbf{Notes} \\
\hline
\hline
Model1: U-Net          & 0.0053 & 0.0532 & 0.0035 & 0.0430 & 15 & 10 & Baseline \\ \hline
\textbf{Model2: RAB-U-Net} & \textbf{0.0036} & \textbf{0.0423} & \textbf{0.0033} & \textbf{0.0416} & 18 & 12 & \textbf{Residual + Attention blocks} \\ \hline
Model3: Res-Net        & 0.0088 & 0.0790 & 0.0086 & 0.0777 & 25 & 15 & Deep residual network \\ \hline
Model4: X-Net          & 0.0168 & 0.1067 & 0.0036 & 0.0440 & 20 & 13 & Extended U-Net variant \\ \hline
\end{tabular}
\end{sidewaystable}

Table~\ref{tab:comparison} compares the performance of four different models based on validation loss (Val\_loss) and validation mean absolute error (Val\_mae). These metrics evaluate the models’ effectiveness and generalization ability for the denoising. Model 2, which incorporates structural modifications to the standard RAB-U-Net, exhibited the best model performance, with the lowest validation loss (0.0036) and validation MAE (0.0423). This indicates the improved efficiency and accuracy of the enhanced architecture.
Figure~\ref{fig:loss_unet}- ~\ref{fig:loss_xnet} depict the training loss and validation loss curves for all four networks throughout training. These graphs provide valuable information about the learning process and the stability of the models:

\begin{figure}[htbp]
\centering
\begin{minipage}{0.48\textwidth}
\centering
\includegraphics[width=\textwidth]{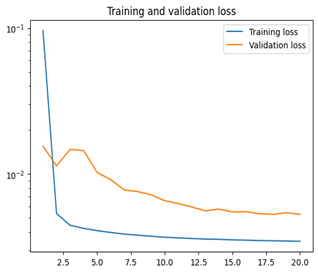}
\caption{Error diagram in U-Net network}
\label{fig:loss_unet}
\end{minipage}
\hfill
\begin{minipage}{0.48\textwidth}
\centering
\includegraphics[width=\textwidth]{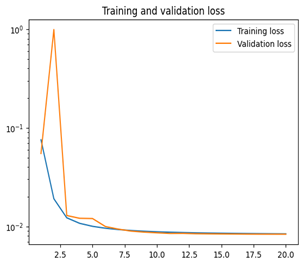}
\caption{Error diagram in RAB-U-Net network}
\label{fig:loss_rabunet}
\end{minipage}

\bigskip

\begin{minipage}{0.48\textwidth}
\centering
\includegraphics[width=\textwidth]{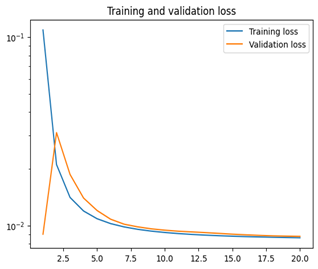}
\caption{Error diagram in Res-Net network}
\label{fig:loss_resnet}
\end{minipage}
\hfill
\begin{minipage}{0.48\textwidth}
\centering
\includegraphics[width=\textwidth]{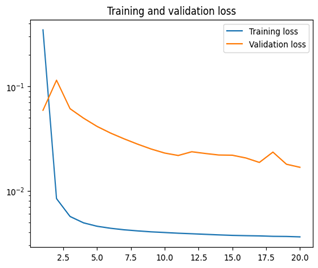}
\caption{Error diagram in X-Net network}
\label{fig:loss_xnet}
\end{minipage}
\end{figure}

\begin{figure}[htbp]
\centering
\begin{minipage}{0.48\textwidth}
\centering
\includegraphics[width=\textwidth]{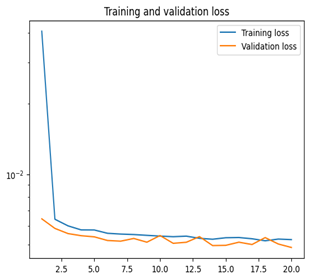}
\caption{Error diagram in U-Net network}
\label{fig:loss_unet1}
\end{minipage}
\hfill
\begin{minipage}{0.48\textwidth}
\centering
\includegraphics[width=\textwidth]{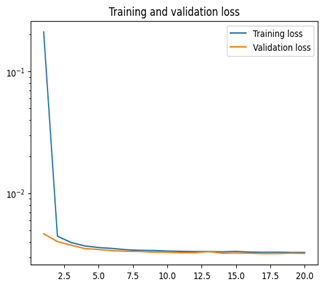}
\caption{Error diagram in RAB-U-Net network}
\label{fig:loss_rabunet1}
\end{minipage}

\bigskip

\begin{minipage}{0.48\textwidth}
\centering
\includegraphics[width=\textwidth]{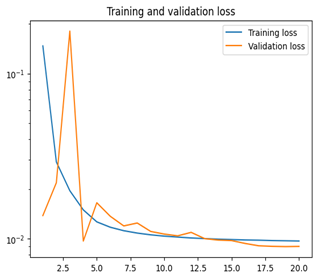}
\caption{Error diagram in Res-Net network}
\label{fig:loss_resnet1}
\end{minipage}
\hfill
\begin{minipage}{0.48\textwidth}
\centering
\includegraphics[width=\textwidth]{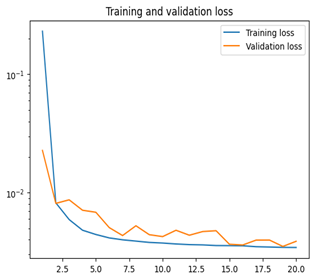}
\caption{Error diagram in X-Net network}
\label{fig:loss_xnet1}
\end{minipage}
\end{figure}

Figure~\ref{fig:loss_unet}- ~\ref{fig:loss_xnet} illustrate the training and validation loss curves for the models used in 20 epochs. As observed, the training loss (blue curve) decreases rapidly during the early epochs, indicating that the model effectively learns the underlying data distribution. The validation loss (orange curve) follows a similar trend, indicating that the models generalize well to unseen data. However, the best result is for the RAB-U-Net. The two curves’ convergence and stabilization towards a small value in Figure 8 show that the model avoids overfitting and achieves a balance between training and validation performance. Minor fluctuations in validation loss are expected due to changes in the validation set. This result confirms the effectiveness of the RAB-U-Net architecture for noise removal. 
The results demonstrate that the RAB-U-Net outperforms all other models, validating its suitability and design enhancements for the problem.
The results obtained using these neural networks for predicting the clean sound of the engine are presented in Figure~\ref{fig:loss_unet1}- ~\ref{fig:loss_xnet1} and Table~\ref{tab:comparison1}. 

\begin{sidewaystable}[htbp]
\centering
\caption{The Val\_Loss and Val\_MAE results for the networks used are presented}
\label{tab:comparison1}
\small
\renewcommand{\arraystretch}{1.08}
\begin{tabular}{|l||c|c|c|c|c|c|l|}
\hline
\textbf{Name of model} & \textbf{Val\_Loss} & \textbf{Val\_MAE} & \textbf{Train-Loss} & \textbf{Train-MAE} & \textbf{Params (M)} & \textbf{Inference Time (ms)} & \textbf{Notes} \\
\hline
\hline
Model1: U-Net          & 0.0049 & 0.0519 & 0.0032 & 0.0398 & 15 &  9 & Baseline \\ \hline
\textbf{Model2: RAB-U-Net} & \textbf{0.0032} & \textbf{0.0417} & \textbf{0.0029} & \textbf{0.0410} & 18 & 11 & \textbf{Residual + Attention blocks} \\ \hline
Model3: Res-Net        & 0.0090 & 0.0774 & 0.0087 & 0.0753 & 25 & 15 & Deep residual network \\ \hline
Model4: X-Net          & 0.0082 & 0.0766 & 0.0032 & 0.0513 & 20 & 14 & Extended U-Net variant \\ \hline
\end{tabular}
\end{sidewaystable}

As explained in the previous section, in this section, too, the best result is related to RAB-U-Net.
Subsequently, based on the accuracy obtained from the two types of data loading, it is concluded that data loading type (b) has performed the best. 
To further substantiate the evaluation of denoising performance, beyond the primary error metrics previously reported, we employ a set of advanced audio quality measures. These metrics provide a more comprehensive assessment by capturing perceptual aspects of audio fidelity and noise suppression effectiveness.
Table~\ref{tab:quantitative} presents the quantitative comparison of Signal-to-Noise Ratio (SNR), Scale-Invariant Signal-to-Distortion Ratio (SI-SDR), and Log-Spectral Distance (LSD) across all investigated models and classical denoising methods.

\begin{table}[htbp]
\centering
\caption{Quantitative comparison of denoising performance using objective metrics}
\label{tab:quantitative}
\small
\renewcommand{\arraystretch}{1.05}
\begin{tabular}{|l||c|c|c|l|}
\hline
\textbf{Model Name} & \textbf{SNR (dB) $\uparrow$} & \textbf{SI-SDR (dB) $\uparrow$} & \textbf{LSD $\downarrow$} & \textbf{Notes} \\
\hline
\hline
Model1: U-Net          & 12.5 & 11.8 & 3.4 & Baseline neural network \\ \hline
\textbf{Model2: RAB-U-Net} & \textbf{14.7} & \textbf{14.2} & \textbf{2.8} & \textbf{Residual + Attention blocks} \\ \hline
Model3: Res-Net        & 10.9 & 10.2 & 4.1 & Deep residual network \\ \hline
Model4: X-Net          &  9.3 &  9.0 & 5.0 & Extended U-Net variant \\ \hline
Band-pass Filter       &  7.5 &  6.9 & 6.5 & Classical deterministic filter \\ \hline
Wavelet Denoising      &  8.2 &  7.7 & 5.9 & Time-frequency thresholding \\ \hline
Wiener Filter          &  7.8 &  7.1 & 6.2 & Adaptive spectral filter \\ \hline
\end{tabular}
\end{table}

Therefore, we proceed with frequency analysis according to the clean sound prediction; according to Table~\ref{tab:peak_frequencies}, we will examine the frequency analysis before and after noise removal.

\begin{table}[htbp]
\centering
\caption{Peak frequencies at three engine speeds in noisy and noise-free conditions}
\label{tab:peak_frequencies}
\footnotesize
\renewcommand{\arraystretch}{1.02}
\begin{tabular}{|c||c|c|c||c|c|c||c|c|c|}
\hline
\multirow{2}{*}{\textbf{}} & 
\multicolumn{3}{c||}{\textbf{1300 Rpm (60s)}} & 
\multicolumn{3}{c||}{\textbf{1750 Rpm (150s)}} & 
\multicolumn{3}{c|}{\textbf{2200 Rpm (300s)}} \\
\cline{2-10}
 & $1f_p$ & $2f_p$ & $3f_p$ & $1f_p$ & $2f_p$ & $3f_p$ & $1f_p$ & $2f_p$ & $3f_p$ \\
\hline
\hline
1 & 31  & 78  & 156 & 63  & 125 & 125 & 78  & 141 & 203 \\ \hline
2 & 78  & 141 & 256 & 78  & 172 & 172 & 94  & 172 & 266 \\ \hline
  & 31  & 78  & 141 & 63  & 125 & 125 & 78  & 141 & 219 \\ \hline
3 & 78  & 109 & 203 & 47  & 94  & 94  & 78  & 156 & 250 \\ \hline
  & 31  & 78  & 141 & 63  & 109 & 109 & 78  & 141 & 203 \\ \hline
4 & 47  & 94  & 141 & 63  & 125 & 125 & 78  & 172 & 250 \\ \hline
  & 31  & 78  & 141 & 63  & 125 & 125 & 78  & 141 & 172 \\ \hline
5 & 78  & 94  & 172 & 63  & 125 & 125 & 78  & 250 & 391 \\ \hline
 & 31  & 78  & 141 & 63  & 109 & 109 & 78  & 141 & 250 \\ \hline
\end{tabular}
\end{table}

\begin{figure}[htbp]
\centering
\includegraphics[width=0.75\linewidth]{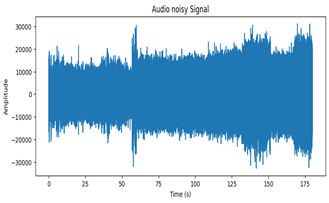}   
\caption{Noise signal}
\label{fig:hot_test3}
\end{figure}

\begin{figure}[htbp]
\centering
\includegraphics[width=0.75\linewidth]{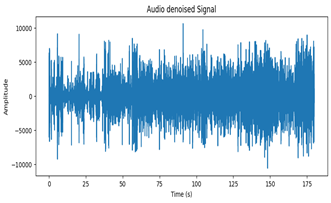}   
\caption{Standardized denoise signal}
\label{fig:hot_test2}
\end{figure}

\begin{figure}[htbp]
\centering
\includegraphics[width=0.75\linewidth]{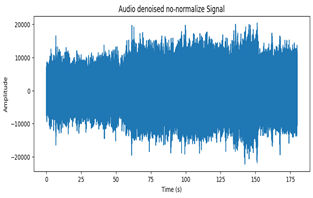}   
\caption{Non-standardized denoise signal}
\label{fig:hot_test1}
\end{figure}

Based on Table~\ref{tab:peak_freq}, which illustrates the peak frequencies, in Table~\ref{tab:peak_frequencies}, a new row has been added to each existing row from the second to the last row. This new row represents the peak frequencies after noise removal. As observed, after noise removal, the frequencies of the first, second, and third peaks at each engine revolution match the frequency of the clean signal with high accuracy. As a result, it can be stated that our network has effectively removed the environmental noise.
During the conversion of the audio from the frequency domain to the time domain after noise removal, some parts of the signal are lost, resulting in unclear engine sound in the noise-free state in the time domain. Therefore, after testing various methods and examining parameters with different values, we concluded that noise removal should be performed on the data without standardization.
This was done using the same method (U-Net and RAB-U-Net neural network) and testing different scenarios. This resulted in hearing the engine sound after noise removal and conversion back to the time domain.
This comparative analysis highlights the importance of network architecture in achieving effective environmental noise removal. The RAB-U-Net emerges as the best-performing model, balancing low error metrics with stable training dynamics. Future work will focus on refining these architectures and exploring additional metrics to comprehensively evaluate their performance.
Next, we will observe the noisy signals, the standardized denoise signal, and the non-standardized noise-free signal in Figure~\ref{fig:hot_test3}- ~\ref{fig:hot_test1}.

Based on the results shown in the figure above and the recorded and heard sounds, we observe that when the data is not normalized, the engine noise, with the ambient noise removed, is easily audible.

\subsection{Summary of Results}

A comparative analysis between classical signal-processing techniques and neural-network models reveals clear differences in their denoising capabilities. Classical methods such as band-pass filtering, wavelet thresholding, and Wiener filtering rely on predefined rules, linear assumptions, or statistical models. As a result, they perform reasonably well for stationary or semi-stationary noise but face limitations when dealing with complex or highly non-linear noise patterns. This is evident in their higher MSE and MAE values, which indicate their inability to fully capture and suppress real-world environmental noise.

In contrast, neural-network-based methods demonstrate significantly improved performance due to their ability to learn non-linear mappings between noisy and clean signals. Models like U-Net and Res-Net already surpass classical techniques by leveraging hierarchical feature extraction and deep representations. The RAB-U-Net further enhances performance by integrating residual connections and attention blocks, allowing the model to focus on relevant features while preserving important signal components. These capabilities lead to lower validation errors and more stable convergence during training.

Moreover, while classical methods require manual parameter tuning and cannot adapt to new noise characteristics, deep-learning models automatically learn noise structures directly from data. This adaptability makes neural networks especially suitable for practical environments where noise conditions vary over time. Therefore, the overall comparison strongly supports the superiority of neural approaches over classical techniques for high-fidelity audio denoising.

\section{Conclusion}
\label{sec:conclusion}

In this paper, we addressed the challenging problem of environmental noise contamination during engine hot tests in real-world production lines. To enable accurate acoustic-based fault diagnosis, a novel deep neural network termed Residual Attention Block U-Net (RAB-U-Net) was proposed for robust background noise removal. The key contribution lies in the synergistic integration of residual feature extraction blocks and attention enhancement modules within the U-Net architecture. These modifications significantly improve feature representation, stabilize gradient flow in deep networks, and enable selective focus on diagnostically relevant engine harmonics while effectively suppressing non-stationary factory noise.

Extensive experiments on real engine recordings demonstrate that the proposed RAB-U-Net consistently outperforms both classical signal processing methods (band-pass filtering, wavelet denoising, Wiener filtering) and contemporary deep learning baselines (vanilla U-Net, Res-Net, X-Net) across all evaluated metrics. Notably, RAB-U-Net achieved the lowest validation MAE of 0.0417 and delivered superior SNR (14.7\,dB), SI-SDR (14.2\,dB), and LSD (2.8), confirming its effectiveness in preserving critical acoustic features essential for downstream fault detection tasks. The combination of residual connections and attention mechanisms proved particularly beneficial in industrial settings, yielding faster convergence, reduced overfitting, and enhanced generalization. With only a modest increase in parameters and inference time, RAB-U-Net offers a practical, deployable solution for real-time engine health monitoring in noisy manufacturing environments.

Future work will focus on lightweight variants for edge deployment, extension to multi-channel microphone arrays, and integration with automated anomaly detection systems to enable fully unsupervised quality control on production lines.

\end{document}